\documentstyle{article}
\def\textbf#1{{\bf #1}}
\def\textit#1{{\it #1}}
\def\text#1{{\rm #1}}
\def\mathrm#1{#1}

\begin{document}

\title{On damage spreading transitions}
\author{Franco Bagnoli\thanks{%
Dipartimento di Matematica Applicata, Universit\`a di Firenze, via S. Marta,
3 I-50139, Firenze, Italy; INFN and INFM sez. di Firenze. 
e-mail:bagnoli@ing.unifi.it}}

\maketitle

\abstract{
We study the damage spreading transition in a generic one-dimensional
stochastic cellular automata
with two inputs (Domany-Kinzel model)
Using an original formalism for the description of the microscopic
dynamics of the model, we are able to show analitically that
the evolution of the damage between two systems driven by the same noise has
the same structure of a directed percolation problem. 
By means of a mean field approximation, we map the density phase
transition into the damage phase transition, obtaining a reliable phase
diagram. We extend this analysis to all symmetric cellular automata
with two inputs, including the Ising model with heath-bath dynamics.}

{\vspace{.5 cm} {\bf Key Words}: Damage spreading, directed percolation,
stochastic cellular automata, disordered systems, symmetry breaking.}

{\vspace{.5 cm} Submitted to J. Stat. Phys. }

{\vspace{.5 cm}}


\section{Introduction}

In this paper we deal with the problem of the evolution of two replicas of a
Boolean system (cellular automaton) that evolve stochastically under the
same realization of the noise. The system is defined on a regular lattice of 
$L$ sites and evolves in discrete time steps. We limit the explicit analysis
to one dimensional systems, but the results can be extended to higher
dimensions.

Let us indicate the time with the index $t=1,\dots ,\infty $ and the space
with $i=0,1,\dots ,L-1$. The state variables $\sigma (i,t)$ can assume the
values 0 or 1 (Boolean variables). The evolution of $\sigma (i,t)$ is given
by probabilistic transition rules and depends on a small number of
neigboring sites; in its simpler form, $\sigma (i,t)$ depends only on the
state of the two nearest neigbors. In this case one can consider the
space-time lattice as a tilted square lattice.

In order to simplify the notation, we write $\sigma _{+}=\sigma (i+1,t)$, $%
\sigma _{-}=\sigma (i-1,t)$, $\sigma ^{\prime }=\sigma (i,t+1)$. The
evolution rule can be synthetically written as 
\[
\sigma ^{\prime }=f(\sigma _{-},\sigma _{+}).  
\]

Since the number of possible values of the couple $(\sigma _{-},\sigma _{+})$
is four, the function $f$ is usually specified by giving the four transition
probabilities $\tau (\sigma _{-},\sigma _{+}\rightarrow 1)$ from each
possible configuration to one:

\begin{equation}
\begin{tabular}{lll}
$\tau (0,0\rightarrow 1)$&$=$ & $p_0$ \\ 
$\tau (0,1\rightarrow 1)$&$=$ & $p_1$ \\ 
$\tau (1,0\rightarrow 1)$&$=$ & $p_2$ \\ 
$\tau (1,1\rightarrow 1)$&$=$ & $p_3$%
\end{tabular}
\label{transprob}
\end{equation}
The normalization condition gives $\tau (\sigma _{-},\sigma _{+}\rightarrow
0)=1-\tau (\sigma _{-},\sigma _{+}\rightarrow 1)$. 

All sites of the lattice are generally updated synchronously. 
Except for the case of deterministic cellular automata, for which the
transition probabilities are either zero or one, we do not expect strong
differences between parallel and sequential updating.

This schematization naturally arise in the modelization of several systems
(contact processes), in physical and biological investigations. It has been
introduced by E. Domany and W. Kinzel\thinspace \cite{domany:DK,kinzel:DK},
and can be considered the prototype for all local stochastic processes. For
a short review of the applicability of this model, see references~\cite
{grassberger:damage} and \cite{dickman:numerical}.

In the thermodynamical limit, the Domany-Kinzel (DK) model exhibits a phase
transition from an ordered to a disordered phase for $p_0=0$. The ordered
configuration is $\sigma (i)=0$ for all $i$ (adsorbing state). The order
parameter is the asymptotic density $m=\lim_{t\rightarrow \infty
}\lim_{L\rightarrow \infty }m(t,L)$, where 
\[
m(t,L)=\frac 1L\sum_{i=0}^{L-1}\sigma (i,t).
\]

In the following, we refer to the critical surface $m=0$ and all its
intersections with planes in the $p_j$ space with the symbol $\alpha $
(see the figures).

This transition has been studied mainly for the symmetric case $p_1=p_2$.
Except for a phenomenological renormalization study to which we collaborated 
\cite{bagnoli:PRG}, the transition line has been found numerically to belong
to the universality class of directed percolation, which is a particular
case of the model. The disagreement for the renormalization group results
can originate from finite-size effects. For the asymmetric case $p_1\neq p_2$%
, it has been claimed \cite{martins:DKnewPhase} that the phase transition
belongs to a different universality class (mean field).

The existence of an adsorbing state is a non-equilibrium feature of the
model, allowing the presence of a phase transition also in a one-dimensional
(spatial) system. It is shown is section\thinspace \ref{section:extensions}
that in the DK model there can be two adsorbing states ($\sigma(i)=0$ and $%
\sigma(i)=1$), related by a simple transformation of the transition
probabilities. The two transition lines met at the point $M$  ($p_1=1/2$, $%
p_3=1$). This point corresponds to the problem of a random walk in one
dimension, and thus exhibit mean field exponents.

A powerful tool for the investigation of this kind of models is the study of
damage spreading. One considers two replicas $\sigma $ and $\eta $ of the
same model, with different initial conditions (they can be completely
uncorrelated or differ only in some sites). The two replicas evolve under
the same realization of the stochasticity. The difference at site $i$ and at
time $t$ between the two configurations is given by 
\[
h(i,t) =\sigma (i,t)\oplus \eta (i,t);
\]
where the symbol $\oplus $ represents the sum modulus two (eXclusive OR,
XOR). Since we use Boolean variables ($a,b\in \{0,1\}$) one can interpret
the exclusive or as $a\oplus b=a+b-2ab$. When mixing XOR and AND
(represented as a multiplication), one can use the algebraic rules for the
sum and the multiplication.

The order parameter for the damage spreading transition is the asymptotic
Hamming distance $H=\lim_{t\rightarrow \infty }\lim_{L\rightarrow \infty
}H(t,L)$ defined as 
\[
H(t,L) =\frac 1L\sum_{i=0}^{L-1}h(i,t), 
\]
using the usual sum.

The critical
surface $H=0$ and its intersection are indicated with the symbol 
$\gamma $.

In the DK model, numerical and analytical investigations \cite
{martins:DKnewPhase,kohring:DKrevisited, zabende:DKphaseDiagram,
rieger:DKreentrant, tome:DKspreading, grassberger:damage} indicated the
existence of a damage spreading phase.

The damage phase transition can be thought as an ergodicity breaking
transition: in the phase where the damage disappears, all initial conditions
asymptotically follow a trajectory that does not depend on the initial
conditions, but only on the realization of the noise. The Hamming distance
can be easily related to the overlap between the configurations. 

The critical exponents for the density and the damage transitions in the
plane $p_1=p_2$ and $p_0=0$ are numerically the same\thinspace \cite
{martins:DKnewPhase,grassberger:damage}. It has been guessed\thinspace \cite
{janssen:guess,grassberger:guess} that all continuous transition from an
adsorbing to and active state belong to the universality class of the DK
model (and thus of directed percolation), and that the same universality
class should include all damage spreading transitions\thinspace \cite
{grassberger:damage}.

Here we want to investigate the connection between the density phase
transition and the damage phase transition in the DK model. We have to
carefully describe the dynamics of the model: the position of the transition
line depends on the way in which the randomness is implemented in the actual
simulations. In section \ref{section:notations} we introduce the formalism
that allows an exact description of how randomness is implemented in the
model. We are thus able to write down the evolution equation for the spins,
and to obtain the evolution equation for the distance between two replicas.
The structure of the latter equation corresponds to the DK model with $p_0=0$%
. We conclude that the universality class of damage spreading is, at least
for this simple case, that of directed percolation. In section \ref
{section:phaseDiagram} we obtain the phase diagram of the DK model by
mapping the transition line for the density to the transition line for the
damage by means of mean field approximations.  In section \ref
{section:extensions} we show that one can infer the existence of a phase
transition for the damage also in cases for which there is no phase
transition for the density, and that there are two disjoint regions in the
parameters space for the damage spreading. Finally, conclusions and open
questions are drawn in the last section.

\section{The damage spreading transition}

\label{section:notations} Let us start from a simple example, the dilution
of rule 90 (in Wolfram's notation\thinspace \cite{wolfram:ca}) that will
also serve to fix the notation. Rule 90 is a deterministic rule that evolves
according with 
\[
\sigma ^{\prime }=\sigma _{-}\oplus \sigma _{+}. 
\]

The transition probabilities for the diluted rule 90 are 
\[
\begin{tabular}{lll}
$\tau (0,0\rightarrow 1)$&$=$&$0$ \\ 
$\tau (0,1\rightarrow 1)$&$=$&$p$ \\ 
$\tau (1,0\rightarrow 1)$&$=$&$p$ \\ 
$\tau (1,1\rightarrow 1)$&$=$&$0,$%
\end{tabular}
\]
$p$ being the control parameter of the model.

In order to apply rule 90 for a fraction $p$ of sites, and rule 0 (all
configurations give 0) for the rest, one usually extracts a random
number $r=r(i,t)$ for each site and at each time step and chooses the
application of rule 90 or rule 0 according with $r<p$ or $r\ge p$ resp.

We can easily write the explicit expression for this rule by means of the
function $[\cdot ]$, assuming that $[\mbox{\it logical proposition\/}]$
takes the value 1 if {\it logical proposition\/} is true, and 0 otherwise
(this interpretation of logical propositions is the standard one in C
language). Finally, we have for the diluted rule 90 
\begin{equation}
\sigma ^{\prime }=[r<p]\left( \sigma _{-}\oplus \sigma _{+}\right) .
\label{rule90}
\end{equation}

One can also think of having all random numbers $r(i,t)$ extracted before
the simulation and attached to the sites of the space-time lattice even
though they are not always used. The random numbers are thus similar to a
space-time quenched (disordered) field.

Once given the set of random numbers, the evolution is completely
deterministic, and the evolution function depends on the lattice position
(spatial and temporal) via the random numbers $r(i,t)$. One can
alternatively define the model stating that deterministic functions are
randomly distributed on the space-time lattice according to a certain
probability distribution. This description is very reminiscent of the
Kauffman model \cite{kauffman:model}.

The damage spreading can be considered a measure of the stability of the set
of possible trajectories, averaging over the realizations of the noise. The
original definition of Lyapunov exponent is a measure of the instantaneous
effects of a vanishing perturbation. Since the state variables of cellular
automata assume only integer values, one has to extend the definition to a
finite initial distance (and to finite time steps), thus taking into account
the possibility of non-linear effects. For cellular automata, the smallest
initial perturbation corresponds to a difference of only one site between
the two replicas. The short-time effects of a (vanishing) perturbation
define the analogous of the derivatives for a continuous system \cite
{bagnoli:derivatives}. The study of the equivalent of the usual (linear)
Lyapunov exponent for deterministic cellular automata allows a
classification of the rules according with the trend of the damage \cite
{bagnoli:lyapunov}. The general problem of damage spreading can thus be
considered equivalent to the study of the non-linear Lyapunov exponent (i.e.
finite initial distance and finite evolution times) for space-time
disordered cellular automata. 

Using the concept of Boolean derivatives \cite{bagnoli:derivatives}, we
develop a Boolean function $f(a,b)$ as 
\[
f(a,b)=f_0\oplus f_1a\oplus f_2b\oplus f_3ab,  
\]
where the Taylor coefficients are 
\begin{eqnarray*}
f_0 &=&f(0,0) \\
f_1 &=&f(0,1)\oplus f(0,0) \\
f_2 &=&f(1,0)\oplus f(0,0) \\
f_3 &=&f(1,1)\oplus f(0,1)\oplus f(1,0)\oplus f(0,0).
\end{eqnarray*}
One can verify the previous expression by enumerating all the possible
values of $a$ and $b$.

Using the bracket $[\cdot ]$ notation, the transition probabilities (\ref
{transprob}) correspond to 
\[
\begin{tabular}{lll}
$f(0,0)$&$=$&$[r_0<p_0]$ \\ 
$f(0,1)$&$=$&$[r_1<p_1]$ \\ 
$f(1,0)$&$=$&$[r_2<p_2]$ \\ 
$f(1,1)$&$=$&$[r_3<p_3],$%
\end{tabular}
\]
where the random numbers $r_j(i,t)$ belongs to the interval $[0,1)$ and
constitute the quenched random field. We neglect to indicate the spatial and
temporal indices for simplicity.

The Taylor coefficients become 
\begin{eqnarray*}
f_0 &=&[r_0<p_0] \\
f_1 &=&[r_1<p_1]\oplus [r_0<p_0] \\
f_2 &=&[r_2<p_2]\oplus [r_0<p_0] \\
f_3 &=&[r_3<p_3]\oplus [r_2<p_2]\oplus [r_1<p_1]\oplus [r_0<p_0].
\end{eqnarray*}

In the following we shall assume $p_1=p_2$ and $r_1=r_2$ so that $f_1 = f_2$
and

\begin{eqnarray*}
f_3 &=& [r_3<p_3]\oplus [r_0<p_0].
\end{eqnarray*}

The correlations among the random numbers $r_j$ (at same space-time
position) affect the position of the critical line for the damage ($\gamma $%
), as already pointed out also by P. Grassberger\thinspace \cite
{grassberger:damage} and E. Domany\thinspace \cite{domany:private}, but not
the position of the transition line for the density ($\alpha $). Only a
careful description of how the randomness is implemented in the model
completely specify the problem of damage spreading. In principle one could
study the case of generic correlations among these random numbers. Here we
consider only two cases: either all $r_j$ are independent (case {\it i},
transition line $\gamma _i$) or they are all identical (case {\it ii},
transition line $\gamma _{ii}$).

The evolution equation for the single site variable $\sigma =\sigma (i,t)$
is 
\begin{eqnarray}
\sigma ^{\prime }&=&[r_0<p_0]\oplus 
\left( [r_1<p_1]\oplus [r_0<p_0]\right)
(\sigma _{-} \oplus \sigma _{+})\oplus  \nonumber \\
&&\left( [r_3<p_3]\oplus [r_0<p_0]\right) \sigma _{-}\sigma _{+}.  \label{DK}
\end{eqnarray}

We can substitute the evolution equation for the replica $\eta =\sigma
\oplus h$, with the evolution equation for the damage $h=\sigma \oplus \eta $%
, obtaining 
\begin{eqnarray}
h^{\prime } &=&\left( [r_1<p_1]\oplus [r_0<p_0]\oplus \left( [r_3<p_3]\oplus
[r_0<p_0]\right) \sigma _{+}\right) h_{-}\oplus  \nonumber \\
&&\left( \lbrack r_1<p_1]\oplus [r_0<p_0]\oplus \left( [r_3<p_3]\oplus
[r_0<p_0]\right) \sigma _{-}\right) h_{+}\oplus  \nonumber \\
&&\left( [r_3<p_3]\oplus [r_0<p_0]\right) h_{-}h_{+}.  \label{DKd}
\end{eqnarray}

This equation has the same structure of the evolution equation of the
original model (\ref{DK}) if in this latter we set $p_0=0$. Remembering that
only for this value of $p_0$ the DK model exhibits a phase transition, we
have a strong argument for the correspondence between directed percolation
and damage spreading transitions. However, also in the symmetric case $%
p_1=p_2$ and $r_1=r_2$, the evolution equation of $h$ is symmetric only in
average, and one has to take into consideration the correlations between $%
\sigma _{-}$ and $\sigma _{+}$. As discussed before, these correlations can
be included in the definition of the DK model, which specify only the
transition probabilities. It remains to be proved that all these versions do
belong to the same universality class.

Let us now consider the case $p_0=0$. Previous numerical investigations
showed that on this plane the two curves $\alpha$ and $\gamma$ meet at the
point $Q = (\sim 0.81,0)$. Inserting the value $p_3=p_0=0$ in the equation (%
\ref{DKd}), we see that the evolution law for $h$ is the same of that for $%
\sigma $, and so both transitions coincide on this line. This corresponds
also to the dilution of rule 90.

Since the rest of the $\gamma $ curve lies away from the density transition
line, the correlations among sites decay rapidly in time and space. This
allow us to use a mean field approximation. In the simplest form, we replace 
$\sigma (i,t)$ with a random bit that assumes the value one with probability 
$m$. With this assumption the equation (\ref{DKd}) becomes 
\begin{eqnarray}
h^{\prime } &=&([r_1<p_1]\oplus [r_3<p_3][r_4<m])h_{-}\oplus  \nonumber \\
&&([r_1<p_1]\oplus [r_3<p_3][r_5<m])h_{+}\oplus [r_3<p_3]h_{-}h_{+};
\label{h}
\end{eqnarray}
where $r_4$ and $r_5$ are independent random numbers. This is a strong
approximation, both because of correlations and because the same $\sigma
(i,t)$ is shared by $h(i-1,t+1)$ and $h(i+1,t+1)$. Nevertheless, we can
assume this equation as a starting point in our derivation of the phase
diagram.

We now want to remap this model onto the original DK model, assuming that
the asymmetry ($r_4\neq r_5$), that in average vanishes, does not strongly
affect the transition.

The remapped transition probabilities $\tilde p$ are
\[
\begin{tabular}{lllll}
$\tilde \tau (0,0\rightarrow 1)$&$=$&$\tilde p_0$&$=$&$0$ \\ 
$\tilde \tau (0,1\rightarrow 1)$&$=$&$\tilde p_1$&$=$&$\pi \left(
[r_1<p_1]\oplus [r_3<p_3][r_5<m]\right)$ \\ 
$\tilde \tau (1,0\rightarrow 1)$&$=$&$\tilde p_1$&$=$&$\pi \left(
[r_1<p_1]\oplus [r_3<p_3][r_4<m]\right)$ \\ 
$\tilde \tau (1,1\rightarrow 1)$&$=$&$\tilde p_3$&$=$&$\pi \left(
[r_3<p_3]([r_4<m]\oplus [r_5<m]\oplus 1)\right),$%
\end{tabular}
\]
where $\pi (f(r))=\int_0^1drf(r)$ is the probability that the Boolean
function $f$ of the random number $r$ takes the value one.

For case {\it i} ($r_1\ne r_3$), we have 
\begin{equation}
\begin{tabular}{lll}
$\tilde p_1$&$=$&$ p_1+p_3m-2p_1p_3m$  \\
$\tilde p_3$&$=$&$p_3(1-2m(1-m)),$
\end{tabular}  
\label{casei}
\end{equation}
while for case {\it ii} ($r_1 = r_3$) 
\begin{equation}
\begin{tabular}{lll}
$\tilde p_1$&$=$&$m|p_1-p_3|+(1-m)p_1;$ \\
$\tilde p_3$&$=$&$p_3(1-2m(1-m)).$  
\end{tabular}
 \label{caseii} 
\end{equation}

Since $\gamma $ lies in the $p_1>p_3$ region, one has for case {\it ii} 
\[
\tilde p_1=p_1-mp_3.  
\]

Notice that for $p_3=0$ or for $m=0$ the two curves $\gamma _i$ and 
$\gamma_{ii}$ coincide, as already noticed numerically by Grassberger \cite
{grassberger:damage}.

Given a certain point $(p_1,p_3)$, it belongs to the damage transition line $%
\gamma $ ($H(p_1,p_3)=0$) if the point $(\tilde p_1,\tilde p_3)$ belongs to
the density transition line $\alpha $ ($m(p_1,p_3)=0$). In order to draw the
phase diagram for the Hamming distance, one has to know the value of the
density $m$ in all the parameter space, and in particular the position of $%
\alpha $. Unfortunately, we do not have a simple expression for these
quantities; in the next section we use some approximation in order to draw a
rough phase diagram. However, we are able to demonstrate that $\alpha $ and $%
\gamma $ are tangent at point $Q$.

The slope $q$ of the normal to $\alpha $ at $Q$ can be given as 
\[
q=\left. \frac{\partial m}{\partial p_1}\right/ \left. \frac{\partial m}{\ {%
\partial p_3}}\right| _Q. 
\]

Considering that $\gamma \equiv H(p_1,p_3)=0\equiv m(\tilde
p_1(p_1,p_3),\tilde p_3(p_1,p_3))=0$, the partial derivatives of $H$ are
given by 
\begin{eqnarray*}
\frac{\partial H}{\partial p_1} &=&\frac{\partial m}{\partial \tilde p_1}%
\frac{\partial \tilde p_1}{\partial p_1}+\frac{\partial m}{\partial \tilde
p_3}\frac{\partial \tilde p_3}{\partial p_1}; \\
\frac{\partial H}{\partial p_3} &=&\frac{\partial m}{\partial \tilde p_1}%
\frac{\partial \tilde p_1}{\partial p_3}+\frac{\partial m}{\partial \tilde
p_3}\frac{\partial \tilde p_3}{\partial p_3}.
\end{eqnarray*}

One has to take into account that $\tilde p_j$ depends on $p_i$ both
directly and via $m$. Inserting the relations \ref{casei} or \ref{caseii}
and considering that at point $Q$, $m=p_3=0$ one obtains that 
\[
q^{\prime }=\left. \frac{\partial H}{\partial p_1}\right/ \left. \frac{%
\partial H}{{\partial p_3}\ }\right| _Q=q. 
\]

Since we know from numerical experiments and from all mean field
approximation beyond the very first one that the slope of $\alpha $ at $Q$
is negative in the $(p_1,p_3)$ plane, the tangency of $\gamma $ to $\alpha $
implies a reentrant behavior for the damage transition curve, as observed in
reference~\cite{rieger:DKreentrant}.

\section{The phase diagram}

\label{section:phaseDiagram} The problem of sketching an approximate phase
diagram for the damage in an analytical way has been dealed with by several
authors \cite{kohring:DKrevisited, rieger:DKreentrant, tome:DKspreading}.
Since any equation for the damage depends on the behavior of one replica,
there are two sources of errors to be controlled: the approximations for the
evolution of one replica and that for the difference (or for the other
replica). As a consequence, all approximation schemes proposed so far
require large efforts for a poor result. Our method is able to exploit the
knowledge of the density phase to study the damage phase transition. There
are several methods that rapidly converge to a good approximation of $\alpha 
$; to our knowledge the best ones are the phenomenological renormalization
group \cite{bagnoli:PRG} and the cluster approximation (local structure) 
\cite{gutowitz:localStructure} improved by finite-size scaling. This latter
method can also give a good approximation of the behavior of $m(p_0,p_1,p_3)$
at any point.

Since here we are not interested in numerical competitions, we use the
high-quality data for the density transition line from reference \cite
{grassberger:damage} combined with a first order mean field approximation
for the density. The $\alpha $ curve has been approximated in the $(p_1,p_3)$
plane by a $5^{\text{th}}$ order polynomial 
\begin{equation}
p_3=\sum_{i=0}^5a_ip_1^i.  \label{alpha}
\end{equation}
The simplest mean field approximation for the asymptotic density $m$ gives 
\[
m=\frac{1-2p_1}{p_3-2p_1}.
\]

By using these approximations one obtains from equations (\ref{casei}) or (%
\ref{caseii}) the curves reported in figure~\ref{figure:phaseDiagram},
together with the presently best numerical results \cite{grassberger:damage}%
.The main source of error is that in the mean field approximation the $\alpha $
curve does not corresponds to the zero of the density. This is particularly
evident in the absence of reentrancy of curves $\gamma _1$ and $\gamma _2$.
Nevertheless even this rough approximation is able to reproduce
qualitatively the phase diagram and to exhibit the influence on the damage
critical line of the different implementations of randomness. Notice that
the $\gamma $ curve from reference \cite{grassberger:damage} corresponds to
the implementation of equation~(\ref{casei}).

\section{The $p_0>0$ case}

\label{section:extensions} The DK model with arbitrary $p_0$ includes all
one dimensional symmetric cellular automaton model or spin system with two
inputs. We can represent each possible model as a point in the three
dimensional unit cube parametrized by $p_0$, $p_1$, $p_3$. The general form
of the transition probabilities from equation (\ref{DKd}) is

\begin{equation}
\begin{tabular}{lll}
$\tilde p_1$&$=$&$p_1+(1-2p_1)(mp_3+(1-m)p_0);$ \\ 
$\tilde p_3$&$=$&$(p_0+p_3-2p_0p_3)(1-2m+2m^2).$%
\end{tabular}
\label{p0g0}
\end{equation}

There is a trivial transformation of the original DK model with $p_0=0$. One
can revert ($0\leftrightarrow 1$) all the spins before and after the
application of the rule. The new transition probabilities $p_i^{\prime }$
are 
\begin{eqnarray*}
p_0^{\prime } &=&1-p_3; \\
p_1^{\prime } &=&1-p_1; \\
p_3^{\prime } &=&1-p_0.
\end{eqnarray*}
The critical plane $p_0=$ $0$ maps to $p_3=1$, and the adsorbing state is
now the configuration in which all spins are one. We indicate with the
symbol $\alpha ^{\prime }$ the critical curve obtained by this
transformation. The point $Q$ is mapped to the point $Q^{\prime }=(1,$ $\sim
0.2,1)$. The parameter cube and the critical curves are reported in figure~%
\ref{figure:cube}. This mapping suggests the presence of a damaged zone near
the corner $(1,0,1)$.

In order to study the position of the critical surfaces for the damage, we
numerically solved equation (\ref{p0g0}) combined with the expression (\ref
{alpha}) of the critical line $\alpha $ in the very simple approximation $%
m=0.5$. The results are reported in figure \ref{figure:cube}. Direct
numerical simulations qualitatively agree with this picture.

The one dimensional Ising model in zero field with heath bath dynamics can
also be expressed with this formalism.

The local field $g=g_i$ for the one dimensional Ising model is 
\[
g=K\left( (2\sigma _{-}-1)+(2\sigma _{+}-1)\right) ,  
\]
where $K=\beta J=J/k_{\text{B}}T$ is the rescaled coupling constant and $%
\sigma =0,1$ the site variables (spin). The local field $g$ can assume the
values $-2K$, $0$, $2K$. 

For the heath bath dynamics, $\sigma'$ takes the value one with probability
 $p$ given by
\[
p=\frac 1{1+\exp (-2g)}.  
\]

The transition probabilities are 
\begin{eqnarray*}
p_0 &=&\frac \xi {1+\xi }; \\
p_1 &=&\frac 12; \\
p_3 &=&\frac 1{1+\xi },
\end{eqnarray*}
where $\xi =\exp (-4K)$. Notice that $p_3=1-p_0$; for $T>0$, $p_0<1/2$,
while for negative temperatures $p_0>1/2$. The point $p_0=p_3=1/2$
corresponds to infinite $T$.

The evolution equation for the site variable is 
\begin{eqnarray*}
\sigma ^{\prime } &=&[r<p_0]\oplus ([r<p_1]\oplus [r<p_0])(\sigma _{-}\oplus
\sigma _{+})\oplus \\
&&([r<p_3]\oplus [r<p_0])\sigma _{-}\sigma _{+};
\end{eqnarray*}
where usually all Taylor coefficients depend on the same random number $%
r=r(i,t)$. The existence line $\omega _{+}$ for the Ising model with $T>0$, $%
p_1=1/2$, $p_3=1-p_0$, intersects the critical line $\alpha $ at $%
M=(0,1/2,1) $. The existence line $\omega _{-}$ for $T<0$ ends at $M^{\prime
}=(1,1/2,0)$. The point $R=(1/2,1/2,1/2)$ corresponds to $T=\infty $ (see
figure \ref{figure:cube}).

The evolution equation for the Hamming distance $h$ is equivalent to
equation (\ref{DKd}), with all $r_j$ equal to $r$. Taking into account the
correlations induced by the random numbers, and that the magnetization is $%
1/2$ except at the critical point, one obtains 
\begin{eqnarray*}
\tilde p_1 &=&\frac{1-\xi }{2(1+\xi )}; \\
\tilde p_3 &=&\frac{1-\xi }{1+\xi },
\end{eqnarray*}
i.e. the line $\chi \equiv p_3=2p_2$, $p_0=0$ for positive or negative
temperatures. The line $\chi $ intersects $\alpha $ at point $M$ for $%
T=0^{\pm }$, confirming that the symmetry breaking transition for the Ising
model occurs at zero temperature.

\section{Conclusions and perspectives}

In this work we presented a formalism that allows the careful description of
Boolean algorithms for stochastic cellular automata (including spin system
like the Ising model). Using this formalism we were able to derive the exact
equation for the evolution of a damage between two replicas that evolve
under the same realization of the noise. Using a mean field hypothesis, we
gave strong indications that the critical line for the damage phase
transition belongs to the same universality class of that for the density in
the DK model, and thus to the directed percolation universality class. We
mapped the density critical line to the damage critical line, obtaining the
regions in the parameter space of a general symmetric cellular automaton
where the replica symmetry breaking is to be expected. Our predictions are
qualitatively confirmed by numerical simulations.

Several questions remain to be answered. Among others: is it possible to
obtain similar results starting from a field description? How does the phase
diagram for more general (asymmetric, three-input, etc.) cellular automata
look like?

\subparagraph{Acknowledgments}

We want to acknowledge fruitful discussions with 
P. Grassberger, R. Livi,  J. Parrondo, 
R. Rechtman and S. Ruffo. We thank P. Grassberger for having
gracefully furnished his high-quality data for the curve $\alpha $ and $%
\gamma $. This work was started during a stage in Mexico with the support of
CNR and CONACYT. Part of the work was performed in Germany with the support
of {\em programma Vigoni} of the {\em Conferenza permanente dei rettori
delle universit\`a italiane}. Finally, the work was terminated during the
workshop {\em Chaos and Complexity} as ISI-Villa Gualino under the CE
contract n. ERBCHBGCT930295.

\newpage

\section*{Figure Captions}

\begin{enumerate}
\item  \label{figure:phaseDiagram}Phase diagram for the density and the
damage in the DK model ($p_0=0$). The curve labeled $\alpha $ is the density
transition line and the one labeled $\gamma $ is the damage transition line
from reference~\cite{grassberger:damage}; the curves labeled $\gamma _i$ and 
$\gamma _{ii}$ correspond to mean field approximation of equations (\ref
{casei}) and (\ref{caseii}) resp.

\item  \label{figure:cube} The parameter cube for the general symmetric
cellular automata. The dashed curves labeled $\alpha $ and $\alpha ^{\prime }
$ belong to planes $p_0=0$ and $p_3=1$ resp., and correspond to the density
phase transitions. The solid curves correspond to the intersection of the
damage critical surface (shaded) $\gamma $ and $\gamma ^{\prime }$ with the
boundaries of the cube. The dotted-dashed lines labeled $\omega _{+}$ and $%
\omega _{-}$ correspond to the existence line for the Ising model for
positive and negative temperatures, resp. The points labeled $M$ and $%
M^{\prime }$ to the critical points of Ising model at zero temperature, and
the point labeled $R$ to infinite temperature. The dotted line labeled $\chi 
$ corresponds to the damage in the Ising model.
\end{enumerate}

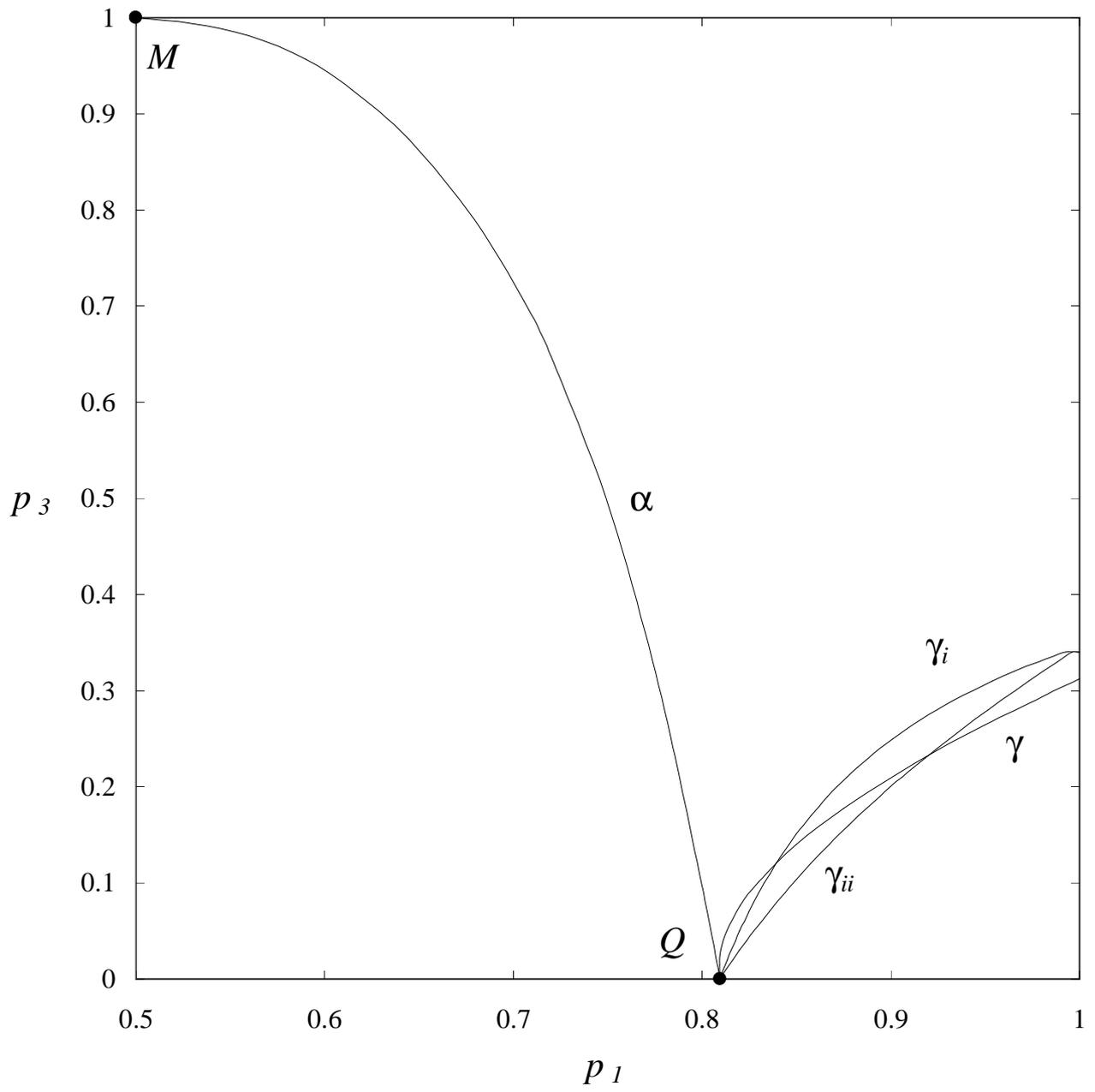
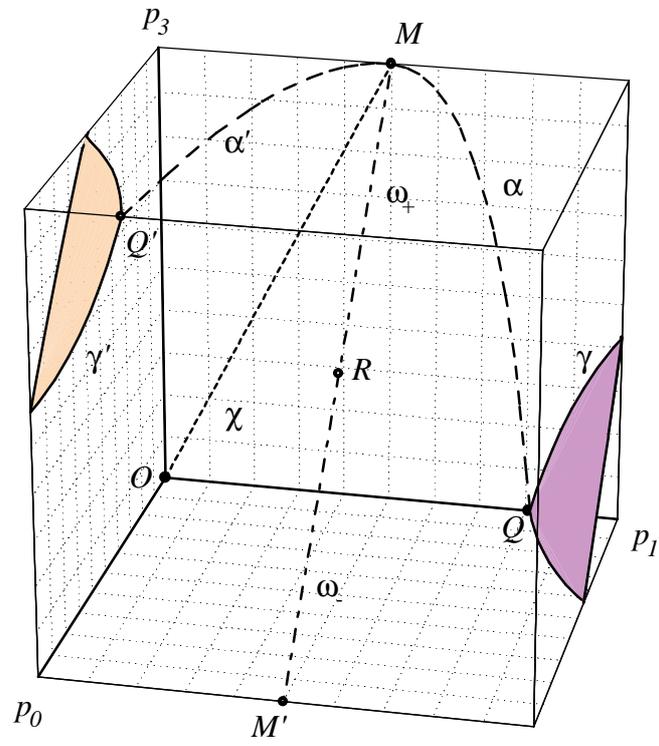
\end{document}